\documentclass[12pt]{ronbun}

\usepackage{color,amsmath,amssymb,graphicx,epsfig}
\usepackage{cite}
\usepackage{bm}
\usepackage{dcolumn}

\newcommand{\nn}{\nonumber}
\newcommand{\e}{{\rm e}}

\newcommand{\del}{\delta}
\newcommand{\ra}{\rangle}

\newcommand{\fr}{\frac}

\numberwithin{equation}{section}

\begin{document}

\begin{flushright}

\parbox{3.2cm}{
{YITP-05-27 \hfill \\ KEK-TH-976 \hfill \\
{\tt hep-th/0506061}}\\
\date
 }
\end{flushright}

\vspace*{0.7cm}

\begin{center}
 \Large\bf Discrete Light-Cone Quantization \\ 
in PP-Wave Background 
\end{center}

\vspace*{3.0cm}

\centerline{\large Kunihito Uzawa$^{\dagger a}$ and   
Kentaroh Yoshida$^{\ast b}$}

\begin{center}
$^{\dagger}$\emph{
Yukawa Institute for Theoretical Physics 
\\ Kyoto University, Kyoto 606-8502, Japan.} \\
$^{\ast}$\emph{Theory Division, 
 High Energy Accelerator Research Organization (KEK),
\\ Tsukuba, Ibaraki 305-0801, Japan.} \\
{\tt E-mail:~$^{a}\,$uzawa@yukawa.kyoto-u.ac.jp \\
\quad  $^b\,$kyoshida@post.kek.jp}\\

\vspace{0.2cm}
\end{center}

\vspace*{1.0cm}

\centerline{\bf Abstract}

We discuss the discrete light-cone quantization (DLCQ) of a scalar field
theory on the maximally supersymmetric pp-wave background in ten
dimensions. It has been shown that the DLCQ can be carried out in the
same way as in the two-dimensional Minkowski spacetime. Then, the vacuum
energy is computed by evaluating the vacuum expectation value of the
light-cone Hamiltonian. The results are consistent with the
effective potential obtained in our previous work [hep-th/0402028].

\vspace*{1.2cm}

\thispagestyle{empty}
\setcounter{page}{0}
\newpage 

\section{Introduction}

Recently, the discrete light-cone quantization (DLCQ) method, which has
been originally developed by \cite{MY,PB}(For reviews, see
\cite{Hiller,Heinzl:2000ht}), has a renewed interest related to the
M-theory formulation \cite{BFSS} (For a review related to the DLCQ, 
see \cite{Banks}). The M-theory formulation \cite{BFSS}
has been extended to the pp-wave background \cite{Blau1,Blau2,Blau3} by
Berenstein-Maldacena-Nastase \cite{BMN}, and so it is also interesting
to study the DLCQ method on the pp-wave background. On the other hand,
for the type IIB string theory on the pp-wave \cite{M,MT}, an
interesting work \cite{Jabbari} has been done related to the DLCQ in the
pp-wave background.

Scalar field theories on pp-wave backgrounds are nice laboratories for
studies of the DLCQ method in the pp-wave case. Furthermore these may give
interesting cosmological models (For studies of cosmological structure
of pp-wave background, see \cite{scalar,stability,PRT,time,SaYo}).   
In fact, scalar field theories on the maximally supersymmetric pp-wave
background in ten dimensions are fairly studied.   
The propagator in this theory was computed in \cite{Mathur}, 
and then the effective potential was discussed in \cite{UY} 
in the path integral formulation \cite{Yan,Heinzl}. 

In this letter, we discuss the DLCQ of scalar field theories in the
maximally supersymmetric pp-wave background in ten dimensions. The
vacuum energies of them are computed by evaluating the vacuum
expectation value (VEV) of the light-cone Hamiltonian. The results
completely agree with the effective potential calculated in our previous
paper \cite{UY}.

\section{DLCQ and Vacuum Energy of a Scalar Field in PP-Wave}

We shall consider a scalar field theory on the maximally supersymmetric 
pp-wave background: 
\begin{eqnarray}
&& ds^2 = 
-2 dx^+dx^- - \sum_{i=1}^8\mu^2(x^i)^2 (dx^+)^2 + 
\sum_{i=1}^8(dx^i)^2\,, \\
&& \qquad F_{+1234} = F_{+5678} = 4\mu\,.
\end{eqnarray}
The constant five-form flux $F$ is a field strength of Ramond-Ramond 
four-form. The light-cone coordinates are defined as 
$x^{\pm} = (x^0 \pm x^9)/\sqrt{2}$\,. 

The action of a scalar field theory on this background is given by 
\begin{eqnarray}
I_{\rm s} &=& \int\!\!d^{10}x\,\sqrt{-g}\left[
-\frac{1}{2}g^{\mu\nu}\partial_{\mu}\phi\partial_{\nu}\phi - V(\phi)
\right] \nn \\
&=& \int\!\!d^{10}x\left[\partial_+\phi\partial_-\phi -
		    \frac{1}{2}\mu^2r^2
\partial_-\phi\partial_-\phi - \frac{1}{2}\partial_i\phi\partial_i\phi 
- V(\phi)\right]\,, 
\label{eq;action}
\end{eqnarray}
where $r^2 = \sum_{i=1}^8(x^i)^2$ and $V(\phi)$ is 
a potential of a scalar field $\phi$\,. 
We will concentrate on the case $V(\phi)=m^2\phi^2/2$ below, 
in order to discuss the DLCQ method for a free scalar field.

The classical equation of motion is 
\begin{eqnarray}
\left(\Box_{\rm (pp)} + m^2\right)\phi = 0\,, \qquad 
\Box_{\rm (pp)} \equiv 2\partial_-\partial_+ - \sum_{i=1}^8\partial_i^2 
- \mu^2 r^2\partial_-^2 \,. 
\label{ceom} 
\end{eqnarray}
By using the identity $x\delta(x)=0$\,, we can easily find the 
solution of (\ref{ceom}): 
\begin{eqnarray}
&& \hspace{-1.5cm}
\phi (x) = \sum_{\{n_i\}}
\int\!\!\frac{dk_-dk_+}{4\pi}\,
\delta(2k_+k_- + |k_-|E_n - m^2) 
\chi_{\{n_i\}}(k_-\,,\;k_+)\,\e^{i (k_+x^+ + k_-x^-)} 
\prod_{i=1}^8\psi_{n_i}(x^i)\,. \label{cs}
\end{eqnarray}
Here we have introduced the following notations:
\begin{eqnarray}
&& E_n \equiv \sum_{i=1}^8 E_{n_i} = \mu\sum_{i=1}^8
\left(n_i + \frac{1}{2}\right) \qquad 
(i=1,\dots,8,\quad n_i=0,\dots,\infty)
\,, \nn \\
&& \psi_{n_i}(x^i) \equiv  N_{n_i}\e^{-\frac{1}{2}\mu |k_-|(x^i)^2}
H_{n_i}(\sqrt{\mu |k_-|}x^i)\,, \qquad N_{n_i} \equiv \left(\frac{\mu
		|k_-|}{\pi}\right)^{1/4}\frac{1}{\sqrt{2^{n_i}
n_i!}}\,. \nn 
\end{eqnarray}
The orthonormal and complete conditions: 
\begin{eqnarray}
\int^{\infty}_{-\infty}\!\!\!\!dx\,\psi_{m}(x)\psi_n(x) = \del_{mn}\,, \qquad 
\sum_{n=0}^{\infty}\psi_n(x)\psi_n(y) = \del(x-y)\,,
\end{eqnarray}
are available to check the normalization of the Poisson bracket. 

After some algebra, the classical solution (\ref{cs}) can be rewritten as 
\begin{eqnarray}
&& \hspace{-1.5cm}
\phi(x) = \frac{1}{4\pi}\sum_{\{n_i\}}\int^{\infty}_{0}\!\!
\frac{dk_-}{k_-}\,\Biggl[
a_{\{n_i\}}(k_-)\,\e^{-i(\hat{k}_+ x^+ + k_- x^-)} 
+ a_{\{n_i\}}^{\ast}(k_-)\,\e^{i(\hat{k}_+ x^+ + k_- x^-)} 
\Biggr]
\prod_{i=1}^8\psi_{n_i}(x^i)\,, 
      \label{eq;pexp}
\end{eqnarray}
where we have defined the on-shell 
energy as $\hat{k}_+ \equiv (|k_-|\,E_n+m^2)/(2k_-)$ 
and redefined the coefficients as follows: 
\begin{eqnarray}
a_{\{n_i\}}(k_-) \equiv
 \chi_{\{n_i\}}(k_-,\,\hat{k}_+)\,, \qquad 
a^{\ast}_{\{n_i\}} (k_-) \equiv \chi_{\{n_i\}}(-k_-\,,\;-\hat{k}_+)\,.  
\end{eqnarray}

The classical Poisson bracket at the simultaneous light-cone time 
$x^{+}=y^{+}$ is given by 
\begin{eqnarray}
\{\phi(x)\,,\;\pi(y)\}_{\rm P} = \frac{1}{2}
\delta(x^- - y^-)\delta^{(8)}(x^i-y^i)\,, \label{eq;pcom}
\end{eqnarray} 
where the $\pi(y)$ is the light-cone canonical momentum: 
\begin{eqnarray}
\pi \equiv \frac{\partial \mathcal{L}}{\partial (\partial_+\phi)} =
 \partial_-\phi\,. 
  \label{eq;pmomentum}
\end{eqnarray} 
Here the factor 1/2 in the r.\ h.\ s.\ of (\ref{eq;pcom}) 
is determined by the Schwinger's action principle and it depends on the 
convention of the light-cone coordinates\footnote{If we take the
light-cone coordinates as $x^{\pm} = x^0 \pm x^{D-1}$\,, then the factor
1/2 is removed.} (For the detail, see \cite{Heinzl:2000ht}). 

In terms of the coefficients $a$ and $a^{\ast}$\,, the 
classical Poisson bracket (\ref{eq;pcom}) is described as 
\begin{eqnarray}
\{a_{\{n_i\}}(k_-)\,,\;a_{\{n'_i\}}^{\ast}(k'_-)
\}_{\rm P} = -4\pi i k_- \delta(k_- - k_-')\prod_{i=1}^8\delta_{n_in_i'}\,.
\end{eqnarray}
It is possible to obtain the usual commutation relation after an
appropriate rescaling as we will see later. 

In the next we will consider to quantize the theory in the canonical
formulation.

\subsection{DLCQ Method in PP-Wave Background}

Now let us consider the canonical quantization of classical scalar field
theories by replacing the classical Poisson bracket of scalar field
$\phi(x)$ and the light-cone canonical momentum $\pi(y)$ with the
commutator at the simultaneous light-cone time $x^{+}=y^{+}$ (For
detail, see \cite{Heinzl:2000ht})\,.  However, we should be careful for
the constraint conditions before the replacement of the commutator.  In
order to carry out the canonical quantization in the light-cone frame,
we follow the DLCQ procedure as a standard manner. In this method, the
physical system is enclosed in a finite volume with a periodic 
boundary conditions in $x^-$ and the longitudinal momentum $k_-$ is 
discretized. 

At first, we compactify the light-cone coordinate $x^-$ as   
$-L\le x^-\le L\,,$ and impose periodic boundary condition for the field;
\begin{eqnarray}
\phi(x^+\,,\;x^-=-L\,,\;x^i)=\phi(x^+\,,\;x^-=L\,,\;x^i)\,.
\end{eqnarray}
The corresponding longitudinal momentum is thus discretized as
$k_{-}=\pi q/L$~($q\in\mathbb{Z}$) and then the Fourier expansion of the
field becomes
\begin{eqnarray}
%  \phi(x^+,\,x^-,\,x^i)
\phi(x) = 
a_{0}(x^+) + 
\sum_{\{n_i\}}\sum_{q>0}\,\frac{1}{\sqrt{4\pi q}}
     \left\{a_{\{n_i\}\,q}(x^+)\;{\rm e}^{-i\frac{\pi q }{L}x^-}
     +a_{\{n_i\}\,q}^{\ast}(x^+)\;{\rm e}^{i\frac{\pi
      q}{L}x^-}\right\} \prod_{i=1}^{8}\,\psi_{n_i,q}(x^i)\,.
       \label{eq;d-expansion}
\end{eqnarray}
Here after the descritization as 
\begin{eqnarray} 
\frac{1}{4\pi} \int_0^{\infty}\!\!\frac{dk_-}{k_-} 
\longrightarrow \frac{1}{4\pi}\sum_{q>0}\frac{1}{q}\,, \qquad 
k_- \del(k_- - k'_-)
\longrightarrow 
q\del_{q,q'}
\,, \nn
\end{eqnarray}
the annihilation and creation operators have been redefined by
rescaling as 
\[
a_{\{n_i\}\,q} \longrightarrow \sqrt{4\pi q}\,a_{\{n_i\}\,q}\,, 
\] 
and the plane-wave with $x^+$ is included in the definition of $a$ 
and $a^{\ast}$\,. 
We also have written the zero mode
$a_{0}$ separately. Though one may think that the zero-mode naively
vanish because of the normalization factors of the Hermite polynomials,
the $k_-=0$ implies a plane-wave expansion rather than a harmonic
oscillator one. Hence we still need to treat carefully the zero-mode
part.  Substituting the expansion (\ref{eq;d-expansion}) into the
Lagrangian (\ref{eq;action}),
 we obtain the following expression (discarding a total time derivative)
\begin{eqnarray}
{\cal L}\left[a_{0}\,,\;a_{\{n_i\}\,q}\right]
  &=& \sum_{\{n_i\}}\sum_{q>0}\left\{
-i a_{\{n_i\}\,q}\:\dot{a}_{\{n_i\}\,q}^{\ast}
- \frac{L}{2\pi q}  \left(\frac{\pi q}{L}\:E_n+m^2\right)
  a_{\{n_i\}\,q}\:a_{\{n_i\}\,q}^{\ast}
\right\}
-m^2 L\:a_{0}^2 
\nn\\
 &=& -i\sum_{\{n_i\}}\sum_{q>0}
    \;a_{\{n_i\}\,q}\:\dot{a}_{\{n_i\}\,q}^{\ast}
-H \,,
\end{eqnarray}
where the Hamiltonian $H$ is given by 
\begin{eqnarray}
H=m^2\:L\:a_{0}^2 + \sum_{\{n_i\}}\sum_{q>0}\;\frac{L}{2\pi q}\:
 \left(\frac{\pi q}{L}\:E_n+m^2\right)
   a_{\{n_i\}\,q}\:a_{\{n_i\}\,q}^{\ast}\,,
 \label{eq;hamilton}
\end{eqnarray}
and the symbol ``$\cdot$'' implies the derivative with respect to
``$x^+$'', such as $\dot{a}_{\{n_i\}\,q}\equiv
 d a_{\{n_i\}\,q}/d x^+$\,. 
From the above, the light-cone Lagrangian is linear with respect to the
velocity i.e., a first order system. In this case we can determine 
the Poisson bracket by comparing the Euler-Lagrange equation and 
the canonical equation as discussed in \cite{Heinzl:2000ht}. 

On the one hand, the Euler-Lagrange equations are 
\begin{eqnarray}
 -i\dot{a}_{\{n_i\}\,q} + \frac{L}{2\pi q}
\;\left(\frac{\pi q}{L}\:E_n+m^2\right)\:a_{\{n_i\}\,q}&=&0\,, 
\label{EL-eq}
\\ 
2m^2L\:a_{0} &=& 0\,. 
\label{eq;constraint}
\end{eqnarray}
The second equation (\ref{eq;constraint}) is non-dynamical and 
gives a constraint implying the absence of zero
mode, i.e. $a_{0}=0$. 

On the other hand, using the Hamiltonian (\ref{eq;hamilton}), 
the canonical equations are 
\begin{eqnarray}
\hspace{-0.9cm}
 \dot{a}_{\{n_i\}\,q} &=& \left\{a_{\{n_i\}\,q}\,,\;H\right\}_{\rm P}\nn\\
       \hspace{-0.9cm} &=& \sum_{\{n_i'\}}\sum_{l>0}\frac{L}{2\pi l}
       \;\left(\frac{\pi l}{L}\:E_n+m^2\right)\,a_{\{n_i'\}\,l}
\{a_{\{n_i\}\,q}\,,\;a^{\ast}_{\{n_i'\}\,l}\}_{\rm P}
       \,.
\end{eqnarray}
These are identical with the Euler-Lagrange equations 
(\ref{EL-eq}) if we identify the canonical bracket as follows: 
\begin{eqnarray} 
\{a_{\{n_i\}\,q}\,,\; a^{\ast}_{\{n_i'\}\,l}\}_{\rm P} = -i\delta_{ql}\cdot 
\prod_{i=1}^8\delta_{n_i,n_i'}\,.
\end{eqnarray}
The constraint condition (\ref{eq;constraint}) can be also derived 
by differentiating the Hamiltonian as 
\begin{eqnarray}
\frac{\partial H}{\partial a_0} = 2m^2 L a_0 = 0\,. 
\end{eqnarray}

The quantization of a free scalar field
is now performed as usual by replacing the classical Poisson bracket 
with the commutator as $[\hat{A}\,,\;\hat{B}]
=i\left\{A\,,\;B\right\}_{\rm P}$\,. 
As the result, the commutation relation of the creation and annihilation
operators is given by
\begin{eqnarray}
[\hat{a}_{\{n_i\}\,q}\,,\;\hat{a}_{\{n_i'\}\,l}^{\dag}]= \delta_{ql}\cdot 
\prod_{i=1}^8\delta_{n_i,n_i'}\,.
   \label{eq:co-re}
\end{eqnarray} 
Finally, the Fock space expansion of the scalar field $\phi$ becomes 
\begin{equation}
\phi\left(x^+\,,\;x^-\,,\;x^i\right)
  =\sum_{\{n_i\}}\,\sum_{q>0}\frac{1}{\sqrt{4\pi q}}
  \left\{\hat{a}_{\{n_i\}\,q}(x^+)\:{\rm e}^{-i\frac{\pi q}{L}x^-}+
   \hat{a}_{\{n_i\}\,q}^{\dag}(x^+)\;{\rm e}^{i\frac{\pi q}{L}x^-}\right\}
\prod_{i=1}^{8}\,\psi_{n_i,q}(x^i)\,. 
 \label{eq;pf}
\end{equation}
As one can see from the expansion (\ref{eq;pf}), the Fock space contains
only particles with positive longitudinal momentum. Operators with
negative longitudinal momentum are annihilation operators. If
longitudinal momentum is conserved, positive and discrete, then states 
with $k_-=q\pi/L$ can have at most $q$ particles among them. Thus, in the
sector with $q$ units of momentum, the theory reduces to
non-relativistic quantum mechanics including eight harmonic
oscillators, with a fixed number of particles. 

Finally, it should be remarked that the structure of Fock space in the
case of pp-wave background is quite similar to that in two-dimensional
Minkowski spacetime. Namely, it is the product of the Fock space of
two-dimensional Minkowski and Hermite polynomials.  From this result,
one can guess that this structure may be extended to other pp-wave
cases.  In the standard pp-wave cases, it is only the difference that
the Hermite polynomials are modified in terms of the oscillation numbers
or the number of harmonic oscillators.  Remarkably speaking, this
structure may be expected to the pp-waves which lead to harmonic
oscillators with negative mass terms.  We encounter this type of
background when Penrose limits of black hole geometries are taken.  In
these pp-waves, hyperbolic cylinder functions, which imply an
instability of the system, would appear instead of Hermite polynomials.
However, the hyperbolic cylinder functions do not mean instabilities as
already discussed in \cite{stability}. On the other hand, our result is
also compatible with this fact because the Fock vacuum
structure is determined by the light-cone directions only. Namely, it is
the two-dimensional Minkowski one. 

In the next subsection we will evaluate the vacuum energy of a scalar
field by using this operator representation.

\subsection{Computation of the Vacuum Energy} 

Next we shall compute the vacuum energy of a free scalar field in
ten-dimensional pp-wave background.  In the previous subsection we have
discussed the DLCQ of a scalar field and derived the operator expression
of the field. We shall compute the vacuum energy density by using this  
quantized scalar field. 

For the Fock vacuum $|0\ra$ of the light-cone Hamiltonian defined by 
$a(k_-)|0\rangle=0$,  
the vacuum energy density $\mathcal{E}$ 
is given by\footnote{For computations of 
vacuum energy of light-cone Hamiltonian in some other models such as 
Gross-Neveu, $SU(N)$ Thirring, $O(N)$ vector models with large $N$
limit, see \cite{KSS}.}
\begin{eqnarray}
\mathcal{E} = \frac{1}{V_9}\langle 0|P_+|0 \rangle\,, \qquad V_9 : 
\mbox{9-dim.\ volume}\,,
\end{eqnarray}
where the light-cone Hamiltonian is 
\begin{eqnarray}
P_+ =\int^{\infty}_{-\infty}\!\!\!\!dx^-\!\!\int^{\infty}_{-\infty}\!\!\!\!
d^{8}x\,
\Bigl\{\pi \partial_{-}\phi-{\cal L}[\phi\,,\;\partial_{\mu}
\phi]\Bigr\}\,. 
 \label{eq:rh} 
\end{eqnarray}
If we use the commutation relation (\ref{eq:co-re}) and 
the expression (\ref{eq;pf})\,, 
the light-cone Hamiltonian is rewritten by 
\begin{eqnarray}
P_+ &=&\sum_{\{n_i\}}\,\sum_{q>0}\,\fr{L}{4\pi q}
       \left(\fr{\pi q}{L}\,E_n+m^2\right)\,
       (2\,\hat{a}_{\{n_i\}\,q}^{\dag}\,\hat{a}_{\{n_i\}\,q}+1) \,.  
\end{eqnarray}
We can evaluate the vacuum energy density $\mathcal{E}$ as 
\begin{eqnarray}
\mathcal{E} &=& \fr{1}{V_9}\,\sum_{n=0}^{\infty}\,{}_{n+7}C_7
                \sum_{q>0}\,\fr{L}{4\pi\,q}
                \left\{\fr{\pi q}{L}\,|\mu|\,(n+4)+m^2\right\}\nn\\ 
            &=& \frac{b+1}{8\pi V_8}\,m^2 \sum_{q>0} \frac{1}{q}\,, 
\end{eqnarray}
where we have used the $V_9 = 2L\cdot V_8$\,, the combination factor 
${}_{n+7}C_7$ denotes the degeneracy of the sum of $n_i$\,, and 
the coefficient $b+1$ is given by
\begin{eqnarray}
b + 1 =\frac{1}{7!}
\left\{\zeta_{\rm R}(-7)-14\zeta_{\rm R}(-5)+49\zeta_{\rm R}(-3)
      -36\zeta_{\rm R}(-1)\right\}=\frac{2497}{3628800}\,. 
\end{eqnarray}
Here $\zeta_{\rm R}(s)$ is the Riemann's zeta function. 
Since the coefficient $b+1$ is
computed by the zeta function regularization, there may be the
ambiguity for the method of mode sum of $n$\,.  When the
continuum limit is considered by taking $L\rightarrow \infty$\,, 
we can recover the expression:
\begin{eqnarray}
\mathcal{E} 
= \frac{b+1}{8\pi V_8}\,m^2 \int^{\infty}_0\!\!dk_-\,\frac{1}{k_-}\,. 
\end{eqnarray}
This can be evaluated by introducing the cut-off 
for the longitudinal momentum as $m^2/\Lambda \leq
|k_-| \leq \Lambda$\,. 
The resulting vacuum energy density is 
\begin{eqnarray}
\label{energy:result}
\mathcal{E} = -\frac{b + 1}{8\pi V_8}\,m^2\,
    \ln\left(\frac{m^2}{\Lambda^2}\right)\,, 
\end{eqnarray}
where the factor $b+1$ denotes the quantum correction which comes from the
 regularization of mode sum of $n$\,. Note that the contribution of the factor
$b+1$ does not appear in the tree level
calculation\cite{Bak:2002ku,Sadri:2003pr}.   
This result agrees with the 1-loop effective potential with $V(\phi)
= m^2\phi^2/2$ that is obtained by using the path integral method
\cite{UY}\,.

On the other hand, we may consider the limit $L\rightarrow \infty$
before the concrete computation, and
introduce the cut-off for the longitudinal momentum as $m^2/\Lambda \leq
|k_-| \leq \Lambda$\,. The $k_-=0$ seems to have a subtlety, but this mode
decouple from the theory as already discussed and so 
this point may have no problem.
After taking the decompactification limit, we use the 
the expansion of scalar field (\ref{eq;pexp}) with 
the creation and annihilation operators whose commutation relations are 
\begin{eqnarray}
[\hat{a}_{\{n_i\}}(k_-)\,,
\;\hat{a}_{\{n'_i\}}^{\dagger}(k'_-)] 
= 4\pi  k_- \delta(k_- - k_-')\prod_{i=1}^8\delta_{n_in_i'}\,.
 \label{eq:cr}
\end{eqnarray} 
Thus, the light-cone Hamiltonian 
$P_+$ is given by using the expressions (\ref{eq;pexp}) and 
(\ref{eq:cr})
\begin{eqnarray}
P_+ &=&\frac{1}{2}\delta(0)
\sum_{\{n_i\}}\int^{\Lambda}_{m^2/\Lambda}\frac{dk_-}{
k_-}\left\{\mu\;k_-\sum^8_{i=1}
\left(n_i+\frac{1}{2}\right)+m^2\right\}\,. 
\label{eq;hamiltonian}
\end{eqnarray}
By using the relation 
$\delta(0)/V_9 = (2\pi)V_1/V_9 = (2\pi)/V_8$\,,  
we can rederive the vacuum energy density (\ref{energy:result}) again. 

 Notably, the effective potential is independent of the parameter $\mu$\,. 
We can see this fact by noting that 
the effect of $\mu$ (i.e., $E_n$) 
can be absorbed by shifting the light-cone momentum
$k_+$\,. But physically, as we discussed in \cite{UY},    
the vacuum energies produced by the quantum 
fluctuations flow from transverse space to the $k_+$-direction 
due to the flux equipped with the pp-wave geometry. 
As the result of the energy flow, the effect of the pp-wave background 
would be realized only as the numerical coefficients, and thus 
the vacuum energy may not explicitly depend on the parameter $\mu$.

\section{A Conclusion and Discussions}

We have discussed the DLCQ of a scalar field theory in the maximally
supersymmetric pp-wave background in ten dimensions.  The DLCQ method in
the pp-wave background has been shown to work well in the same way as in
two-dimensional case, even if we consider higher dimensional theory. It
is because transverse momenta are discretized and so the treatment in the
pp-wave case is quite similar to the two-dimensional Minkowski case. 

We also calculated the vacuum energy of a free scalar field in the
pp-wave and it has been shown the resulting vacuum energy surely agrees
with the effective potential obtained in our previous paper.

The DLCQ method may play an important role in the M-theory formulation
\cite{BFSS}, where the discrete light-cone quantized M-theory can be
described by a matrix model. It would be an important key ingredient 
in studies of the pp-wave matrix model \cite{BMN}. 
We believe that our study of DLCQ in a scalar field theory on the
pp-wave background should be a clue to shed light on some features of
DLCQ method in the pp-wave case.

\section*{Acknowledgments}

This work of K.~U.\ is partially supported by Yukawa fellowship. 
The work of K.~Y.\ is supported in part by JSPS Research Fellowships for
Young Scientists.

\end{document}